# MoS$_2$ Transistors Operating at Gigahertz Frequencies

Daria Krasnozhon, Dominik Lembke, Clemens Nyffeler, Yusuf Leblebici,

Andras Kis[*]

Electrical Engineering Institute, Ecole Polytechnique Federale de Lausanne (EPFL), CH-1015 Lausanne, Switzerland

*Correspondence should be addressed to: Andras Kis, andras.kis@epfl.ch

The presence of a direct band gap[1-4] and an ultrathin form factor[5] has caused a considerable interest in two-dimensional (2D) semiconductors from the transition metal dichalcogenides (TMD) family with molybdenum disulphide (MoS$_2$) being the most studied representative of this family of materials. While diverse electronic elements[6,7], logic circuits[8,9] and optoelectronic devices[12,13] have been demonstrated using ultrathin MoS$_2$, very little is known about their performance at high frequencies where commercial devices are expected to function. Here, we report on top-gated MoS$_2$ transistors operating in the gigahertz range of frequencies. Our devices show cutoff frequencies reaching 6 GHz. The presence of a band gap also gives rise to current saturation,[10] allowing power and voltage gain, all in the gigahertz range. This shows that MoS$_2$ could be an interesting material for realizing high-speed amplifiers and logic circuits with device scaling expected to result in further improvement of performance. Our work represents the first step in the realization of high-frequency analog and digital circuits based on two-dimensional semiconductors.

The operating frequency of semiconductor devices has increased remarkably since the invention of the first transistor. By reducing the transistor's critical dimensions, we can at the same time reduce the electron transit time and the gate capacitance which boosts the performance of transistors in the high-frequency range. The next challenging step is to reduce the device dimensions in the vertical direction to their ultimate limit. For this purpose, nanoscale materials such as carbon nanotubes (CNTs)[11,12] and graphene[13-15] have been suggested. Theory predicts that CNTs operating in the ballistic regime could provide a gain in the THz range.[16] However, their small size carries disadvantages, such as the large contact resistance resulting in high device impedance, making it difficult to measure and integrate CNT-based devices with a standard radio-frequency (RF) setup.

Another candidate for future high-frequency devices is graphene. Graphene shows large carrier mobility[17] and saturation velocity[18]. Graphene field-effect transistors (GFETs) have lower impedance compared to CNT-FETs, making impedance matching a less sensitive issue in GFETs than in CNT-FETs. The other important advantage of graphene compared to CNTs lies in its 2D nature and high mobility which results in large transconductance $g_m$ and intrinsic cutoff frequency $f_T$,[13,19,20] the frequency at which current gain becomes unity.

Power and voltage amplification are also important transistor applications in electronic circuits. The maximum frequency of oscillation $f_{max}$ is the frequency at which the power gain equals unity. In high-performance voltage and power amplifiers, $f_{max}$ should be as high as possible. In the case of RF-GFETs, $f_{max}$ still remains lower than the current gain cutoff frequency $f_T$[19,20] because it is relatively difficult to achieve saturation in graphene FETs due to the lack of band gap. In devices that lack saturation in drain current, this results in high values of drain conductance $g_{ds}$ and limited voltage gain $A_v = g_m/g_{ds}$.

Other two-dimensional materials such as for example molybdenum disulphide (MoS$_2$) could be interesting for applications in RF electronics, especially in high-frequency digital electronics where voltage gain higher than 1 is desired. MoS$_2$ is a semiconductor from the family of transition-metal dichalcogenide (TMD) materials with the common formula MX$_2$, where M represents a transition metal (M = Mo,W, Nb, Ta, Ti, Re) and X stands for either Se, S or Te. MoS$_2$ has a bandgap and unique valley[21,22] and spin properties as well as remarkable properties for new electronic and optoelectronic applications.[23] In contrast to graphene which is a semimetal, MoS$_2$ is a two-dimensional semiconductor with an electronic structure dependent on its thickness.[1-4] First locally gated FETs based on monolayers of MoS$_2$[24] were shown to have room-temperature current on/off ratio higher than $10^8$. Monolayer MoS$_2$ also shows voltage gain[25] larger than 10, due to its natural band gap, high intrinsic transconductance and drain-source current saturation.[10] The combination of all these properties makes MoS$_2$ attractive for high-frequency applications. Self-consistent ballistic quantum transport simulations, independent of mobility values, show that cutoff frequencies well above 100 GHz could be possible for MoS$_2$ transistors with channel lengths of $L_g \sim 15$ nm, operating in the ballistic limit.[26] In addition, large-area MoS$_2$ can be prepared using either liquid-phase exfoliation[27] or CVD growth.[28-30]

Here, we characterize top-gated MoS$_2$ FETs in the high-frequency range. The current gain of MoS$_2$ FETs decreases with increasing frequency and shows the typical $1/f$ dependence for different thicknesses of 2D MoS$_2$ crystals. We realized MoS$_2$ FETs with a maximum transconductance $g_m = 54$ µS/µm, the highest transconductance in MoS$_2$ transistors reported up to date[10], and showing current saturation. The highest intrinsic cut-off frequency $f_T$ derived from the measured scattering parameters is 6 GHz and presents a strong dependence on the bias voltage, gate voltage and material thickness. MoS$_2$ RF transistors also show a voltage gain higher than 1 and a maximum frequency of oscillation comparable to the cut-off frequency, reaching $f_{max} = 8.2$ GHz which makes MoS$_2$ interesting for high-frequency logic circuits and amplifiers.

Our MoS$_2$ FETs are fabricated from MoS$_2$ exfoliated onto a highly resistive intrinsic Si substrate covered with 270 nm thick SiO$_2$.[31] Figure 1 shows the device layout of a MoS$_2$ FET with probe pads in the ground-signal-ground configuration (GSG) designed for high-frequency measurements. We have fabricated RF transistors based on 1L-MoS$_2$, 2L-MoS$_2$, 3L-MoS$_2$ and multilayer Figure S1 (5 nm thick) MoS$_2$ crystals shown on Figure 1b. Electrical contacts were patterned using electron-beam lithography and by depositing 90-nm-thick gold electrodes. This was followed by an annealing step at 200°C in order to remove resist residue and decrease the contact resistance. Atomic layer deposition (ALD) was used to deposit a 30 nm thick layer of high-κ dielectric HfO$_2$. Local top gates for controlling the current in the two branches were fabricated using another e-beam lithography step followed by evaporation of 10 nm/50 nm of Cr/Au. All our devices had a gate length $L_g$ = 240 nm with the underlap region 50 nm long on both sides of the gate electrode. Channel widths are in the 9 - 21.5 µm range due to the stochastic nature of the $MoS_2$ exfoliation process.

Transfer and output characteristics of 1L-$MoS_2$ and 3L-$MoS_2$ FETs are shown in Figure 2. All our devices show transconductance typical of n-type semiconductors with on-state current reaching ~300 µA/µm for $V_{ds}$ = 2V and gate voltage $V_{tg}$ = 10 V in the case of monolayer $MoS_2$. The dielectric constant of 30nm $HfO_2$ deposited at 200°C is 14, probably due to surface impurities trapped between the $HfO_2$ and $MoS_2$ layer. From the $I_{ds}$-$V_{tg}$ characteristic or our devices, we extract estimates for the field effect mobility and contact resistance $\mu_{FE}$ = 85 $cm^2$/Vs and $R_c$ = 2.0 kΩ·µm for the case of the monolayer device and $\mu_{FE}$ = 51 $cm^2$/Vs and $R_c$ = 3.1 kΩ·µm for the trilayer device (Fig. S1 in Supplementary information).

One of the most important parameters affecting the high-frequency performance of transistors is the transconductance $g_m = dI_{ds}/dV_{tg}$, shown in the insets of Figure 2a and b. The magnitude of $g_m$ drastically rises when increasing $V_{ds}$ in the range from 100 mV to 1.5 V for 1L-$MoS_2$ and from 100 mV to 2 V for 3L-$MoS_2$, reaching a maximum of ~ 44 µS/µm for the single-layer and ~54 µS/µm for the trilayer device. These values are comparable to those reported for first graphene RF devices[13] and are desirable for achieving high-frequency operation. Our devices can also easily achieve saturation. Figures 2c and d show the $I_{ds}$-$V_{ds}$ characteristics of two $MoS_2$ devices with a different number of layers. We measure very low values for the channel conductance $g_{ds}$, reaching values as low as ~ 3 µS/µm in the trilayer device at $V_{ds}$ = 3V and $V_{tg}$ = 4V.

One of the main figures of merit to estimate the performance of radio-frequency devices is the cut-off frequency $f_T$, the frequency at which the current gain becomes unity[32]:

$$f_T = \frac{g_m}{2\pi} \frac{1}{(C_{gs}+C_{gd})[1+g_{ds}(R_s+R_d)]+C_{gd}g_m(R_s+R_d)}$$

where $g_m$ is the intrinsic transconductance, $C_{gs}$ is the gate-source capacitance, $C_{gd}$ is the gate-drain capacitance, $g_{ds}$ the drain conductance, $R_s$ and $R_d$ the source and drain series resistances, respectively.

To probe the intrinsic performance of $MoS_2$ FETs in the RF range of operation, we measure the scattering S-parameters describing our devices using a vector network analyzer and perform standard calibration using dummy OPEN and SHORT structures[33] with the purpose of de-embedding the influence of the parasitic, gate capacitance and resistance due to contact pads and device connections. These structures were fabricated with the same layout as the $MoS_2$ FETs and produced in the same fabrication run on intrinsic Si substrates, Figure S3. The intrinsic cut-off frequency obtained in this way is related to the carrier transit time between the source and drain terminals. We analyzed FETs based on 1L-$MoS_2$, 2L-$MoS_2$, 3L-$MoS_2$ as well as multilayer $MoS_2$ stacks with thicknesses of ~5 nm. All the devices have a 240 nm gate length, total channel length of 340 nm and a 30 nm thick layer of $HfO_2$ acting as the gate dielectric.

Figure 3 shows the short-circuit current gain (the ratio of small-signal drain and gate currents) $h_{21}$ as a function of frequency for different thicknesses of exfoliated $MoS_2$, before and after de-embedding. We operated our devices under bias and gate voltages $V_{ds}$ and $V_{tg}$ where they showed the highest intrinsic transconductance. We attribute the significant noise in the low-frequency characteristics to the large impedance mismatch between our devices and the test setup. The current gain $h_{21}$ decreases with increasing frequency. All our devices showed $f_T$ higher than 1 GHz after de-embedding. For 1L-MoS$_2$, we obtain a cut-off frequency $f_T$ = 2 GHz for $V_{ds}$ = 2 V and $V_{tg}$ = -3 V. We record the highest cut-off frequency for trilayer MoS$_2$ with $f_T$ = 6 GHz for $V_{ds}$ = 2.5 V and $V_{tg}$ = -1.5 V. To estimate the value of $f_T$ independently, we also used the Gummel's method,[34,35] with results shown on Figure S2. The cut-off frequencies obtained from intercept of the 1/$f$ dependence and Gummel's method are closely matched.

Figure 4 represents the behavior of the cut-off frequency as a function of the number of layers of MoS$_2$ FETs. We see a rise of the cut-off frequency with increasing number of layers from 1L-MoS$_2$ to 3L-MoS$_2$, while the 5 nm thick multilayer MoS$_2$ devices shows no improvement in performance over the trilayer device. Possible reasons for this could include a lower contact resistance in the case of the trilayer device and longer collision-free paths for the trilayer device as indicated by the more pronounced saturation behavior in the trilayer device when compared to the monolayer device.

Since the intrinsic cut-off frequency is mostly determined by the minimum time required for charge carriers to travel across the channel, decreasing the length of the semiconducting channel and the gate length are expected to result in increased cut-off frequency of FETs as reported for RF GFET devices.[13,35-37] Because the cut-off frequency $f_T$ is proportional to transconductance $g_m$ which in turn is proportional to the field-effect mobility $\mu_{FE}$, improving the mobility of the material by removing the adsorbates[38], removing intrinsic defects and reducing the temperature should result in further improvements to the cut-off frequency.

In addition to current gain, transistors for high-frequency logic circuits and amplifiers should also show voltage and power gains characterized by the maximum frequency of oscillation $f_{max}$. This is the frequency at which the power gain is equal to 1. Just as $f_T$, $f_{max}$ can be expressed by following the equation for field-effect transistors:[39]

$$f_{max} = \frac{f_T}{2\sqrt{g_{ds}(R_g + R_s) + 2\pi f_T C_g R_g}}$$

where $g_{ds}$ is the drain differential conductance, $R_s$ the source resistance, $C_{gd}$ the gate to drain capacitance and $R_g$ is the gate resistance which mainly depends on the gate thickness and area.

In figure 5, we plot Mason's unilateral power gain $U$ as a function of frequency for two MoS$_2$ devices. This figure of merit is useful for describing the quality of three-terminal devices and is the highest possible gain obtained by unilateralizing the two-port network with lossless feedback. For active devices, the quantity $U$ is higher than 1 and $f_{max}$ corresponds to the frequency at which $U$ = 1 ($U$ = 0 dB). As shown on Figure 5a, we obtain $f_{max}$ = 2.2 GHz for single-layer MoS$_2$ with 240 nm gate length. For trilayer MoS$_2$, we obtained $f_{max}$ = 8.2 GHz, Figure 5b, similar to high-end RF FETs based on epitaxial graphene.[15] This result shows that MoS$_2$ could be interesting for high-frequency amplifiers. We also extract the intrinsic voltage gain $A_v = g_m/g_{ds}$ for the same set of devices by converting the scattering S-parameters parameters to impedance Z-parameters where $A_v = Z_{21}/Z_{11}$.[39] The results are presented in Figures 5c and d for our MoS$_2$ FETs. At 200 MHz, $A_v$ is equal to 1.9 (5.5 dB) and is twice as high as in state of the art RF-GFETs with equivalent gate oxide thickness and similar gate lengths.[37] Further device scaling is expected to improve all the operating parameters described here. Improvements in the voltage gain can be expected from decreasing the oxide thickness, resulting in better electrostatic control over the semiconducting channel. Reducing the gate length and improving the semiconducting channel mobility is expected to result in increased cut-off frequency $f_T$, while the maximum frequency of oscillation $f_{max}$ can be increased by reducing the parasitic effects due to its layout, capacitance, gate and contact resistance.

In conclusion, we have characterized the high-frequency operation of top-gated MoS$_2$ transistors with a 240 nm gate length. Our MoS$_2$ RF-FETs show an intrinsic transconductance higher than 50 μS/μm, saturation of drain-source current and a voltage gain higher than 1. These features allow the operation of MoS$_2$ transistors in the GHz range of frequencies. Our devices show cut-off frequencies as high as 6 GHz and are able to not only amplify current in this frequency range but also power and voltage, with the maximum frequency of oscillation $f_{max}$ = 8.2 GHz. These features show that MoS$_2$ and other 2D TMD semiconductors can be used to fabricate transistors that operate at gigahertz frequencies. Further improvements in device geometry and material processing are possible and will help realize the full potential of 2D semiconductors for low-cost and flexible high-frequency analog current and voltage amplifiers as well as high-frequency logic circuits.

## MATERIALS AND METHODS

Device characterization is performed at room temperature in an RF probe station (Cascade Microtech). DC and AC voltages are applied using bias tees connected to each probe head. DC currents are measured using an Agilent B2912A parameter analyzer. AC excitations and subsequent S-parameter measurements are performed using an Agilent N5224A vector-network analyzer (VNA). High-frequency S-parameter characterization was carried out in the 0.2-5 GHz frequency range. The two VNA ports are connected to the sample's source and drain electrode via the bias-tees and probe heads. The probe system was calibrated before the measurements of the device under test (DUT) in order to take into account any spurious contribution of connectors, cables and the electrical environment of the DUT and to subtract it from the measured signal. This is done first by means of a calibration pad and in the second step using a set of dummy structures. The system was calibrated using the LRRM (Line – Reflect – Reflect – Match) method for the required frequency range and at low input power (typically -27 dBm) using a standard CSR-8 substrate. On-chip OPEN and SHORT structures with exact design layout of the devices were used to de-embed the parasitic effects such as pad capacitance and interconnection resistance with the purpose to obtain the intrinsic RF performance. OPEN dummy structures are used to cancel the influence of the capacitive coupling between the electrodes. SHORT dummy structures are used to cancel the series resistance from the leads and contact resistance between probe tips and landing pads.

## ACKNOWLEDGEMENTS

The authors would like to thank W. Grabinski and O. Sanchez for useful discussions as well as A. Allain and S. Bertolazzi for technical help. Device fabrication was carried out in the EPFL Center for Micro/Nanotechnology (CMI). We thank Z. Benes (CMI) for technical support with e-beam lithography. This work was financially supported by Swiss SNF grants 135046 and 144985 as well as ERC grant no. 240076.

**SUPPORTING INFORMATION AVAILABLE**

Supplementary figures and discussion related to device fabrication, field-effect mobility and contact resistance extraction, cut-off frequency extraction using the Gummel's method, layout of structures used for parameter de-embedding. This material is available free of charge via the Internet at http://pubs.acs.org.

**FIGURES**

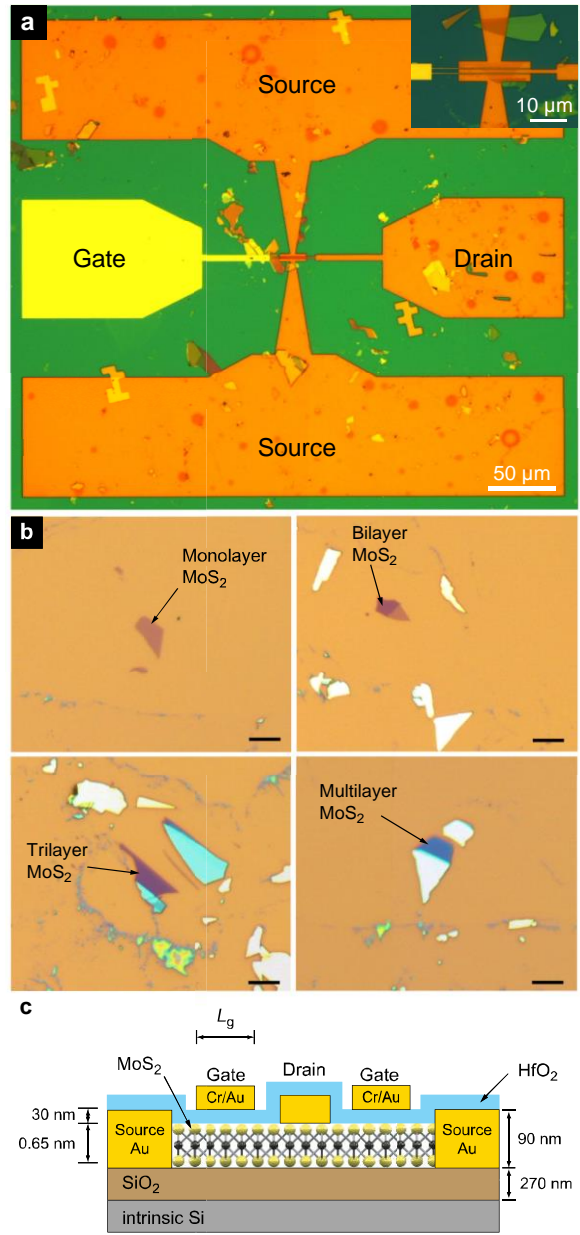

**Figure 1. MoS$_2$-based high-frequency transistors. a,** Optical image of the device layout with ground – signal – ground pads for the drain and the gate on intrinsic Si substrate covered with a 270 nm thick SiO$_2$ layer. The inset shows the magnified optical image of the coplanar waveguide with the gate length $L_g$ = 240 nm. The total channel length is 340 nm. Underlap regions are 50 nm long. **b,** Optical images of 1L-MoS$_2$, 2L-MoS$_2$, 3L-MoS$_2$ and multilayer (5 nm thick) MoS$_2$. Transistors are fabricated on top of regions with uniform crystal thickness. Scale bar is 10 μm long. **c,** Schematic cross-sectional view of the RF-MoS$_2$ transistor.

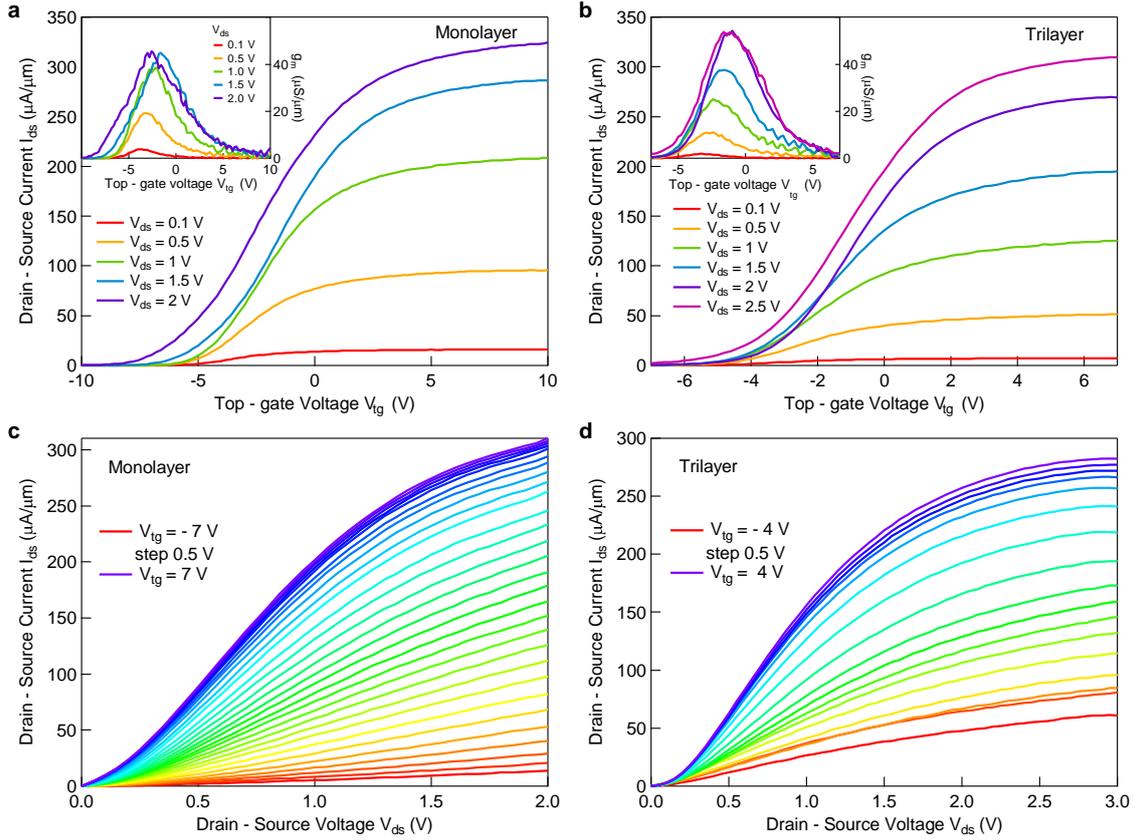

**Figure 2. DC output characteristics of MoS$_2$ RF transistors. a,** Transfer characteristic for 1L-MoS$_2$ FET under the applied drain voltage in the range from $V_{ds}$ = 0.1 V - 2 V with a step of 0.5 V. **b,** Transfer characteristic for 3L-MoS$_2$ FET under the applied drain voltage in the range from $V_{ds}$ = 0.1 V - 2.5 V with a step of 0.5 V. Insets in a and b show the transconductance of 1L-MoS$_2$ and 3L- MoS$_2$ derived from $I_{ds} - V_{tg}$ characteristics. **c,** The drain current $I_{ds}$ as a function of bias voltage $V_{ds}$ measured for different top – gate voltages in the case of a device based on single-layer MoS$_2$. The top-gate voltage varies from -7 V to 7 V with a step of 0.5 V. **d,** The drain current $I_{ds}$ as a function of bias voltage $V_{ds}$ for the 3L-MoS$_2$ device, measured for different top – gate voltages. The gate voltage varies from -4 V to 4 V with a step of 0.5 V.

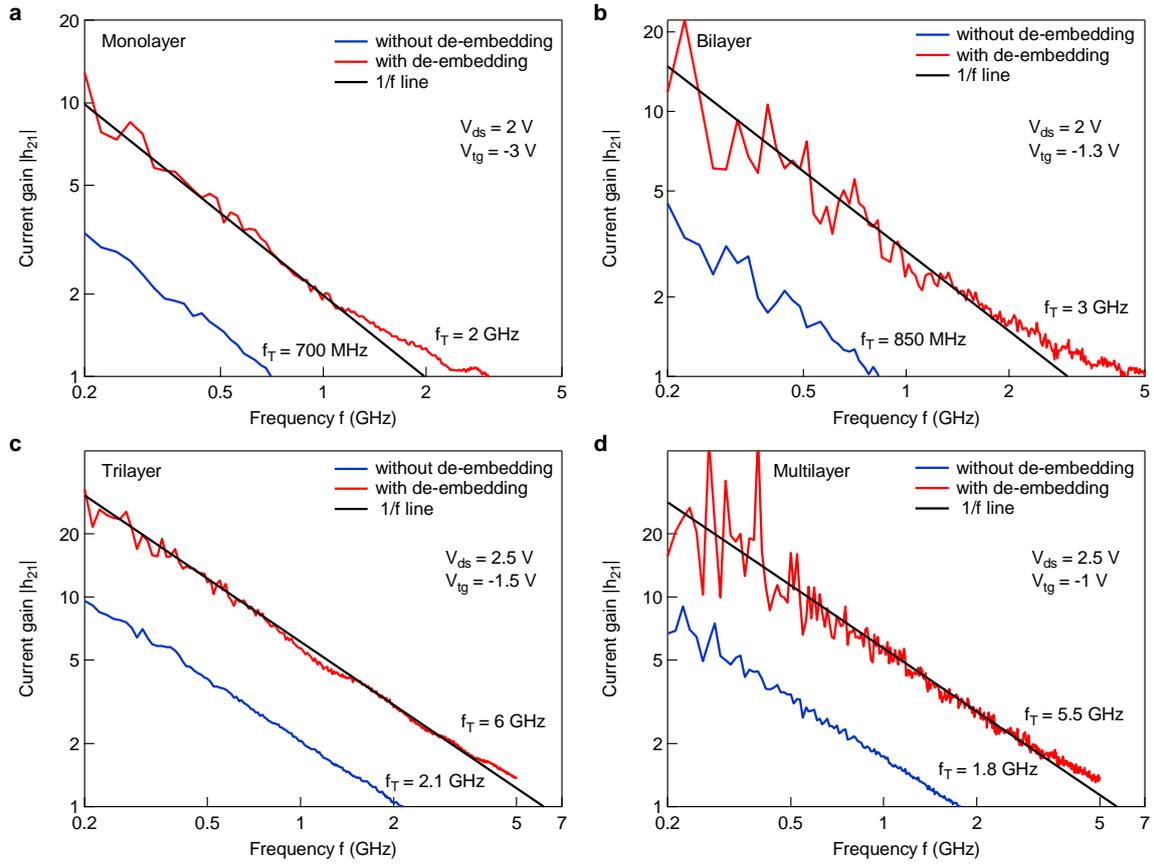

**Figure 3. RF performance analysis for devices based on exfoliated MoS$_2$.** Small-signal current gain $h_{21}$ as a function of frequency for devices based on 2D MoS$_2$ crystals with different thicknesses, with and without deembedding. **a,** Single-layer MoS$_2$ with the cutoff frequency $f_T$ = 2 GHz. **b,** bilayer MoS$_2$ with $f_T$ = 3 GHz at applied $V_{ds}$ = 2 V and $V_{tg}$ = -1.3 V. **c,** Trilayer MoS$_2$ with $f_T$ = 6 GHz and **d,** multilayer MoS$_2$ with $f_T$ = 5.5 GHz at applied $V_{ds}$ = 2.5 V and $V_{tg}$ = -1 V. The current gain $h_{21}$ decreases with increasing frequency, following the -20dB/dec slope expected for conventional FETs.

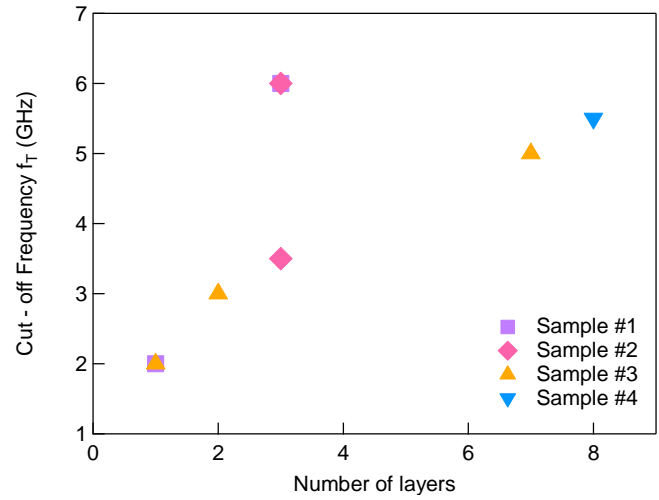

**Figure 4. Summary of the RF performance of MoS$_2$ devices. a,** Intrinsic cut-off frequency $f_T$ as a function of the number of layers of MoS$_2$ transistors for 4 different samples with $L_g$ = 200 nm. Highest cut-off frequency is observed for the trilayer transistor.

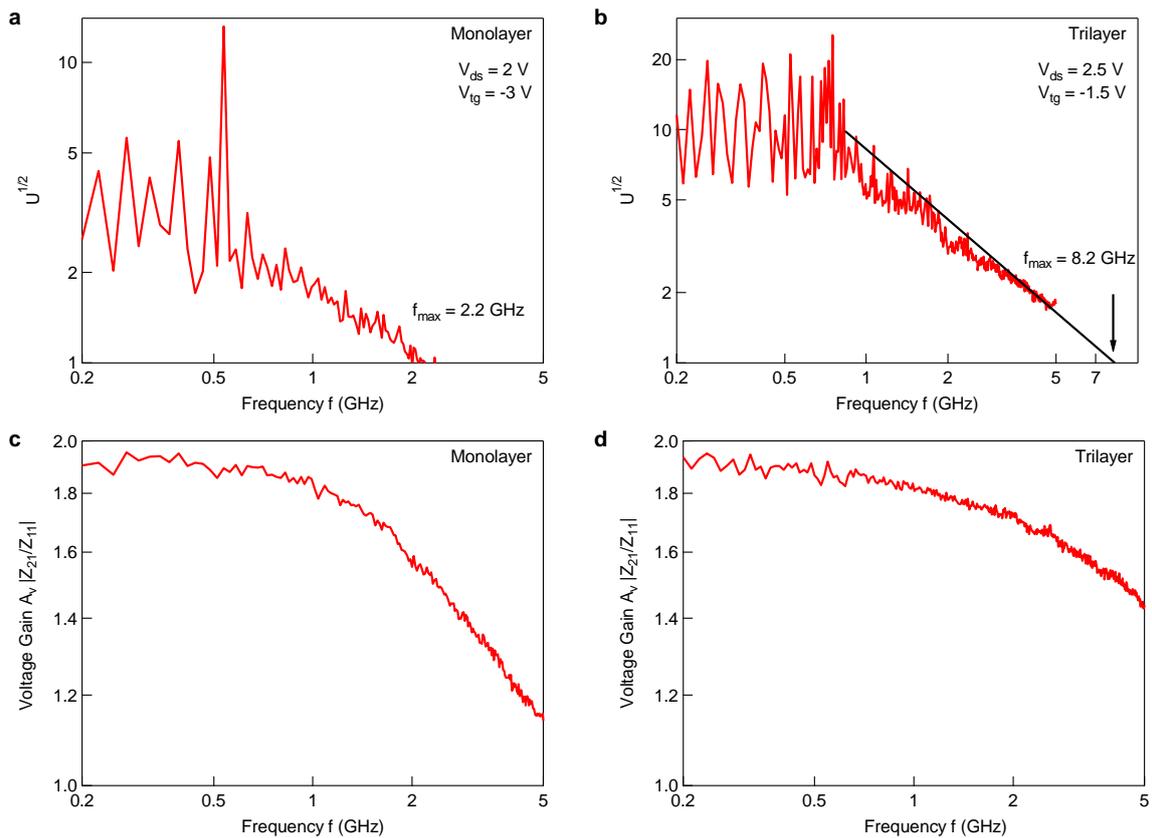

**Figure 5. Power and voltage gain analysis for devices based on exfoliated MoS$_2$. a,** Mason's unilateral gain $U$ as a function of frequency for the monolayer MoS$_2$ device. Maximum frequency of oscillation $f_{max}$ = 2.2 GHz is extracted at the point where power gain reaches unity. **b,** Same as in part a but for a trilayer device, resulting in $f_{max}$ = 8.2 GHz. **c,** Voltage gain $|Z_{21}/Z_{11}|$ as a function of frequency for the monolayer MoS$_2$ device showing gain higher than 1 up to 5 GHz. **d,** Dependence of the voltage gain on frequency for the trilayer device.